\documentclass[12pt]{iopart}

\usepackage{graphicx}

\begin{document}

\title[ ]{Spin-dependent electron transport in a Rashba quantum wire with rough edges}

\author{Xianbo Xiao$^{1,2}$, Huili Li$^{1}$, Guanghui Zhou$^{3}$ and Nianhua Liu$^{2\dag}$
\footnotetext{Author to whom any correspondence should be addressed. } }

\address{$^1$ School of Computer, Jiangxi University of Traditional Chinese Medicine, Nanchang 330004, China.\\
$^2$ Institute for Advanced Study, Nanchang University, Nanchang 330031, China.\\
$^3$ Department of Physics, Hunan Normal University, Changsha 410081, China.}
\ead{nhliu@ncu.edu.cn}

\begin{abstract}
We investigate theoretically the spin-dependent electron transport in a Rashba
quantum wire with rough edges. The charge and spin conductances are calculated
as function of the electron energy or the wire length by adopting the spin-resolved
lattice Green function method. For a single disordered Rashba wire, it is found that
the charge conductance quantization is destroyed by the edge disorder. However, a
nonzero spin conductance can be generated and its amplitude can be manipulated by
the wire length, which is attributed to the broken structure symmetries and the
spin-dependent quantum interference induced by the rough boundaries. For a large
ensemble of disordered Rashba wires, the average charge conductance decreases
monotonically, however, the average spin conductance increases to a maximum
value and then decreases, with increasing wire length. Further study shows
that the influence of the rough edges on the charge and spin conductances
can be eliminated by applying a perpendicular magnetic field to the wire.
In addition, a very large magnitude of the spin conductance can be achieved
when the electron energy lies between the two thresholds of each pair of
subbands. These findings may not only benefit to further apprehend the
transport properties of the Rashba low-dimensional systems but also provide
some theoretical instructions to the application of spintronics devices.

\end{abstract}

\maketitle

\section{Introduction}

In 1990, Datta and Das proposed a novel device, i.e. spin-field-effect
transistor \cite{Datta}, which demonstrates that the spin state of the
conduction electrons in a quasi-two-dimensional electron gas (2DEG) can
be controlled via the Rashba spin-orbit interaction (SOI) \cite{Rashba,Bychkov}.
From then on, the spin-dependent electron transport in low-dimensional
semiconductor quantum systems under the modulation of Rashba SOI has
drawn intensive investigation \cite{Zutic}, since it is the physical
basis of semiconductor spintronics devices. In these devices, the spin
instead of charge degree of freedom of the conduction electrons is utilized
to store and communicate information.

Modern nanofabrication techniques allow the manufacture of various
high quality low-dimensional quantum structures, however, the
disorders such as impurities, defects and rough edges are inevitable
in these systems\cite{Saminadayar}. Therefore, the effects of these
disorders on the spin-dependent electron transport should be
considered in the application of semiconductor spintronics devices.
In previous works, the influence of the scattering caused by the
impurities and defects on the function of the semiconductor
spintronics devices such as spin filter
\cite{Shi,Yamamoto,Ohe,Rebei,Yamamoto2}, spin transistor
\cite{Wang,Shen}, spin diode \cite{Xiao} and spin separator
\cite{Xiao2} has been studied extensively. Furthermore, the impurity
and defect disorders also result in the strength of the SOI to
fluctuate randomly in space, leading to many new phenomena such as
the realization of the minimum possible strength of SOI
\cite{Sherman2} and the localization of the edge electrons for
sufficiently strong electron-electron interactions \cite{Strom}.
However, recent studies show that the scattering induced by the
impurities and defects has the same impact on every propagating
channel in low-dimensional quantum structures while the scattering
caused by the rough edges mainly impact the highest propagating
channel \cite{Garcia1,Gil,Garcia2,Markos,Perez,Feilhauer}. Thereby,
the edge disorder may give rise to some new effects in the
spin-dependent electron transport of the low-dimensional quantum
systems under the modulation of the Rashba SOI. To the best of our
knowledge, the effects of the edge disorder on the spin-dependent
electron transport have not yet been investigated up to now.

In order to uncover the novel transport phenomena induced by the
rough edges in low-dimensional quantum structures, in this paper we
theoretically calculate the charge and spin conductances at low
temperatures for a Rashba quantum wire in the presence of the edge
disorder. It is demonstrated that the spin conductance displays very
different behaviors compared to those of the charge conductance when
spin-unpolarized electrons are injected from the input lead. The
quantum charge conductance steps for a single disordered Rashba wire
are destroyed and decreases dramatically with increasing wire
length, that is, edge disorder strength. However, a spin-polarized
current can be generated in the output lead and its magnitude is
very sensitive to the wire length. Similarly, the average charge
conductance for a large ensemble of disordered Rashba wires are
suppressed gradually while the average spin conductance are enhanced
firstly and then suppressed, with the increasing of the wire length.
Further study shows that the edge disorder effects on the
spin-dependent electron transport can be removed when a
perpendicular magnetic field is applied to the quantum wire. The
quantum charge conductance steps recover one by one with an increase of the magnetic field
strength. However, the spin conductance almost disappears except within the energy window between the
two thresholds of each pair of subbands, where a very large amplitude of spin
conductance can be obtained.

The rest of the paper is organized as follows. In section 2, the
theoretical model and the spin-resolved lattice Green function
approach are presented. Some numerical examples and discussions of
the results are demonstrated in section 3. Finally, section 4
concludes the paper.

\section{Model and Method}
The system studied in present work is schematically depicted in
figure 1, where a 2DEG in the $(x, y)$ plane is restricted to a
quantum wire with rough edges by a confining potential $V(x, y)$.
The 2DEG is confined in an asymmetric quantum well, where the SOI is
assumed to arise dominantly from the Rashba mechanism. The quantum
wire has a length $L$ and a width $W(x)$ randomly fluctuating with
$x$, connected to two ballistic semi-infinite leads of constant
width $W$. The two connecting leads are normal-conductor electrodes
without SOI, since the spin-unpolarized injection is interested.
Such kind of Rashba system can be described by the discrete lattice
model. The spin-resolved tight-binding Hamiltonian including the
Rashba SOI on a two-dimensional lattice is given as follows
\cite{Wang2},
\begin{eqnarray}
H=H_0+H_{so}+V,
\end{eqnarray}
where
\begin{eqnarray}
H_0=\sum\limits_{lm\sigma}\varepsilon_{lm\sigma}c_{lm\sigma}^{\dag}c_{lm\sigma}-t\sum\limits_{lm\sigma}\{c_{l+1m\sigma}^{\dag}c_{lm\sigma}\nonumber\\
+c_{lm+1\sigma}^{\dag}c_{lm\sigma}+H.c\},
\end{eqnarray}
\begin{eqnarray}
H_{so}=t_{so}\sum\limits_{lm\sigma\sigma'}\{c_{l+1m\sigma}^{\dag}(i\sigma_{y})_{\sigma\sigma'}c_{lm\sigma'}\nonumber\\
-c_{lm+1\sigma}^{\dag}(i\sigma_{x})_{\sigma\sigma'}c_{lm\sigma'}+H.c\},
\end{eqnarray}
and
\begin{eqnarray}
V=\sum\limits_{lm\sigma}v_{lm}c_{lm\sigma}^{\dag}c_{lm\sigma},
\end{eqnarray}
in which $c_{lm\sigma}^{\dag}(c_{lm\sigma})$ is the creation
(annihilation) operator of electron at site $(lm)$ with spin
$\sigma$, $\sigma_{x(y)}$ is Pauli matrix, and
$\varepsilon_{lm\sigma}=4t$ is the on-site energy with the hopping
energy $t=\hbar^{2}/2m^{\ast}a^{2}$, here $m^{\ast}$ and $a$ are the
effective mass of electron and lattice constant, respectively.
$v_{lm}$ is the additional confining potential. The SOI strength is
$t_{so}=\alpha/2a$ with the Rashba constant $\alpha$. When a
magnetic field $\vec{B}(0,0,1)$ is introduced, it could be
incorporated into the nearest-neighbor hopping energy by the
Peierl's phase factor such as
\begin{eqnarray}
T_{lm,lm+1}=texp(i\hbar\omega_{c}l/2t)=(T_{lm+1,lm})^{\ast};~~
T_{lm,l+1m}=(T_{l+1m,lm})^{\ast}=t,
\end{eqnarray}
where $\omega_{c}=eB/m^{\ast}c$ is the cyclotron frequency. In order
to keep the transitional symmetry of system along the $x$-axis
direction, the vector potential is chosen as $\vec{A}=(By,0,0)$. The
SOI strength $t_{so}$ should be given the same modification as that
of the hopping energy when the magnetic field is presented.
Moreover, the Zeeman effect from the external magnetic field is not
included here.

In order to calculate the Green function of the whole system through
the recursive Green function method, the tight-binding Hamiltonian
(1) is divided into two parts in the column cell
\begin{eqnarray}
H=\sum\limits_{l\sigma\sigma'}H_{l}^{\sigma\sigma'}+\sum\limits_{l\sigma\sigma'}(H_{l,l+1}^{\sigma\sigma'}+H_{l+1,l}^{\sigma'\sigma}),
\end{eqnarray}
where $H_{l}^{\sigma\sigma'}$ is the Hamiltonian of the $l$th
isolated column cell, $H_{l,l+1}^{\sigma\sigma'}$ and
$H_{l+1,l}^{\sigma'\sigma}$ are intercell Hamiltonian between the
$l$th column cell and the $(l+1)$th column cell with
$H_{l,l+1}^{\sigma\sigma'}=(H_{l+1,l}^{\sigma'\sigma})^\dag$, here
the lattice position parameter $l$ is within the range $[1,N]$. The
Green function of the whole system can be computed by a set of
recursive formulas \cite{Lee2,Ando},
\begin{eqnarray}
\langle l+1|G_{l+1}|l+1\rangle^{-1}=E-H_{l+1}-H_{l+1,l}\langle l|G_{l}|l\rangle H_{l,l+1},
H_{l+1,l}\langle l| G_{l}|0\rangle,
\end{eqnarray}
\begin{eqnarray}
\langle l+1|G_{l+1}|0\rangle=\langle l+1|G_{l+1}|l+1\rangle H_{l+1,l}\langle l| G_{l}|0\rangle,
\end{eqnarray}
in which $\langle l|G_{l}|l\rangle$ and $\langle l|G_{l}|0\rangle$
are respectively the diagonal and off-diagonal Green function, and
\begin{eqnarray}
H_{l+1}=\left(
\begin{array}{cc}
H_{l+1}^{\sigma\sigma} & H_{l+1}^{\sigma\sigma'} \\
H_{l+1}^{\sigma'\sigma} & H_{l+1}^{\sigma'\sigma'} \end{array}
\right),~~ H_{l+1,l}=(H_{l,l+1})^\dag=\left(
\begin{array}{cc}
H_{l+1,l}^{\sigma\sigma} & H_{l+1,l}^{\sigma\sigma'} \\
H_{l+1,l}^{\sigma'\sigma} & H_{l+1,l}^{\sigma'\sigma'} \end{array}
\right).
\end{eqnarray}

The recursion starts from the Green function of the left
semi-infinite lead without SOI $\langle 0|G_{0}|0\rangle$. By
utilizing equation (7), the following sequences of the diagonal
Green function $\langle 0|G_{0}|0\rangle\rightarrow\langle
1|G_{1}|1\rangle\rightarrow\cdots\rightarrow\langle
N|G_{N}|N\rangle$ can be obtained. Substituting these diagonal Green
functions into equation (8), the off-diagonal Green functions
$\langle 1|G_{1}|0\rangle\rightarrow\langle
2|G_{2}|0\rangle\rightarrow\cdots\rightarrow\langle
N|G_{N}|0\rangle$ can be achieved in turn. In the final recursion
step, the Green function of the right semi-infinite with a vanishing
SOI $\langle N+1|G_{N+1}|N+1\rangle$ is attached, and then the Green
function of the whole system $\langle N+1|G_{N+1}|0\rangle$ is
generated finally. Here the Green function of the left and right
semi-infinite leads can be obtained analytically from reference
\cite{MacKinnon}.

Utilizing the Green function of the whole system calculated above,
one can compute the two-terminal spin-resolved conductance through
the Landauer-B$\ddot{u}$ttiker formula \cite{Buttiker}
\begin{eqnarray}
G^{\sigma'\sigma}=e^2/hTr[\Gamma_{L}^{\sigma}G^{r}\Gamma_{R}^{\sigma'}G^{a}],
\end{eqnarray}
where $\Gamma_{L(R)}=i[\sum_{L(R)}^{r}-\sum_{L(R)}^{a}]$ with the
self-energy from the left (right) lead
$\sum_{L(R)}^{r}=(\sum_{L(R)}^{a})^{\ast}$, the trace is over the
spatial and spin degrees of freedom, and $G^{r}(G^{a})$ is the
retarded (advanced) Green function of the whole system and
$G^{a}=(G^{r})^\dag$.

In the following calculations, all the energy and lengths are
normalized by the hopping energy $t(t=1)$ and the lattice constant
$a(a=1)$, respectively. The $z$-axis is chosen as the spin-quantized
axis so that $|\uparrow>=(1,0)^{T}$ represents the spin-up state and
$|\downarrow>=(0,1)^{T}$ denotes the spin-down state, where $T$
means transposition. The width of the two semi-infinite leads is
fixed at $W=20$, the average value of the width of the Rashba
quantum wire with rough boundaries is taken to be $\langle W(x)
\rangle=17$, i.e., $W(x)$ oscillating randomly within the range from
$14$ to $20$. The strength of SOI is set at $t_{so}=0.08$.  For
simplicity, the hard-wall confining potential approximation is
adopted to determine the boundaries of the quantum wire, due to
different confining potentials only alter the positions of the
subbands and the energy gaps between them.

For a single disordered Rashba wire, the charge conductance and the
amplitude of the spin conductance of $z$-component are defined as
$G^e=G^{\uparrow\uparrow}+G^{\uparrow\downarrow}+G^{\downarrow\downarrow}+G^{\downarrow\uparrow}$
and
$G^{Sz}=\frac{e}{4\pi}|\frac{G^{\uparrow\uparrow}+G^{\uparrow\downarrow}-G^{\downarrow\downarrow}-G^{\downarrow\uparrow}}{e^2/h}|$,
respectively. Here the charge conductance means the transfer
probability of electrons, and the spin conductance represents the
change in local spin density between the input lead and the output
lead caused by the transport of spin-polarized electrons
\cite{Khomitsky}. In addition, the equation that is usually used for
calculation of the localization length \cite{Nikolic} is modified to
$\lambda=-1/\lim\limits_{L\rightarrow
\infty}\frac{1}{L}Tr|G^{a}G^{r}|$ as the spin degree of freedom is
included.

For a large number of disordered Rashba wires, the average charge
conductance and the charge conductance fluctuation are respectively
quantified as $G_{a}^{e}=\langle {G^{e}}\rangle$ and
$G_{f}^{e}=[\langle {(G^{e})}^{2}\rangle-\langle G^{e}\rangle
^{2}]^{\frac{1}{2}}$, where $\langle\cdots\rangle$ denotes averaging
over an ensemble of samples with different realizations of edge
disorder \cite{Nikolic2}. Similarly, we define the average spin
conductance and the spin conductance fluctuations as
$G_{a}^{Sz}=\langle {G^{Sz}}\rangle$ and $G_{f}^{Sz}=[\langle
{(G^{Sz})}^{2}\rangle-\langle G^{Sz}\rangle ^{2}]^{\frac{1}{2}}$,
respectively.

\section{Results and discussions}

\subsection{The conductance for a single Rashba quantum wire with rough edges}

Figure 2(a) shows the charge conductance as function of the electron
energy $E$ for various wire lengths. By comparing with the charge
conductance $G^{e}$ of the ideal straight Rashba wire $W(x)\equiv17$
(dotted line), the step structures of the charge conductance are
deteriorated and its magnitude decreases dramatically within the
whole energy band range even for a very small edge disorder strength
$L=10$ (dash-dotted line). The charge conductance curve also shows
sample specific fluctuations, resulting from the quantum
interference effect induced by the rough edges. The typical spacing
between peaks and valleys in the charge conductance depends on the
wire length as $E_{c}\sim 1/L$ \cite{Nikolic2}. Consequently, the
charge conductance fluctuates  quickly as the wire length $L$ is
increased to $40$ (dashed line). When the wire length is increased
further to $200$, plenty of resonant peaks appear in the charge
conductance, with the maximum value $2e^{2}/h$ (solid line). In addition, the charge conductance falls faster near
the right energy band edge than that near the energy band center
though there are more propagating channels, which can be understood
by the localization length $\lambda$ plotted in figure 2(c). Due to
the localization lengths near the band center are larger than those
near the band edges to the right, the charge conductance near the
band center is subjected to less impacts of the edge disorder.
Figure 2(b) plots the amplitude of the spin conductance of
$z$-component as function of the electron energy for the same Rashba
quantum wire with rough edges. Surprisingly, the spin conductance
$G^{Sz}$ exhibits very different transmission behaviors from those
of the charge conductance. Owing to the longitudinally symmetry of
the straight Rashba quantum wire \cite{Zhang}, the amplitude of the
spin conductance keeps zero for the whole energy band (dotted line).
However, for the Rashba wire with boundary roughness ($L=10$), a
nonzero spin conductance is generated (dash-dotted line) and its
maximum amplitude is enlarged when the wire length is increased to
$40$ (dashed line). When the wire length is increased further to the
strong localization regime $L=200$ $(L>\lambda)$, the amplitude of
the spin conductance also shows plenty of resonant peaks, but the
maximum value is smaller than $e^{2}/h$ (solid line). The mechanism
of the edge-disorder-induced spin conductance is attributed to the
broken longitudinal symmetry of the Rashba wire \cite{Xiao,Zhai} and
the spin-dependent quantum interference between the forward
propagating channels and the backward propagating channels caused by
the rough boundaries. Furthermore, due to the scattering by the
rough edges is weakest in the lowest propagating channel and
strongest in the highest propagating channel
\cite{Garcia1,Gil,Garcia2,Markos,Perez,Feilhauer}, the
spin-dependent quantum interference mainly happens in the highest
propagating channel. As a consequence, the amplitude of the spin
conductance is not larger than $e^{2}/h$.

According to figure 2, it is understood that the charge and spin
conductances are very sensitive to the length of the Rashba quantum
wire. In order to clarify the contribution of the wire length, the
charge and spin conductances as function of the wire length are
plotted in figures 3(a) and 3(b), respectively. The electron energy
$E=2.08$. It is found that the sensitivity of the charge and spin
conductances to the wire length differs from that to the electron
energy. As shown in figure 3(a), the charge conductance decreases
sharply with the raising of the wire length if only it is smaller
than the localization length $\lambda=140$ (see figure 2(c)).
However, when the wire length is increased to the strong
localization regime ($140<L\leq300$), the charge conductance is of
order $2e^{2}/h$ and  fluctuates more obviously. As the wire length
is further increased ($L>300$), the charge conductance almost
disappears in virtue of the strong backscattering caused by the edge disorder. In contrast with the charge conductance, the amplitude
of the spin conductance oscillates quickly with an increase of the
wire length, as shown in figure 3(b). There are many peaks and
valleys exist in the spin conductance and the maximum value of these
peaks almost reaches $e^{2}/h$ when the wire length is increased to
the strong localization region. Further, the amplitude of the spin
conductance becomes very small as the wire length $L>300$, since the
charge conductance almost decreases to zero.

\subsection{The average conductance and the conductance fluctuations for a large ensemble of Rashba quantum wires with rough
edges}

Figure 4(a) shows the ensemble average charge conductance as
function of the electron energy for various wire lengths. For the
short Rashba wires $L=10$, the average charge conductance
$G_{a}^{e}$ increases sharply with the raising of the electron
energy except for a short plateau of the first propagating channel,
indicating that the edge disorder has less impacts for the lower
propagating channel (longer wavelength). However, for the longer
Rashba wire, the average charge conductance increases slowly with
increasing electron energy $(L=40)$ or is independent of the
electron energy within a wide range $(L=200)$. In addition, there is
a broad charge conductance peak near the band edge to the left,
which corresponds to peak in the localization length plotted in
figure 2(c). The corresponding ensemble average spin conductance as
function of the electron energy is illustrated in figure 4(b). By
comparing to the average charge conductance, the average spin
conductance $G_{a}^{Sz}$ displays converse behaviors. The average
spin conductance is almost independent of the electron energy near
the band center for the short Rashba wire $L=10$, however, it
depends on the electron energy for the longer Rashba wire $L=40$ and
its magnitude is increased. Especially, when the wire length locates within
the strong localization regime $(L=200)$, the average spin
conductance is very sensitive to the electron energy and behaves
like the localization length shown in figure 2(c). Figure 4(c) shows
the charge conductance fluctuation as function of the electron
energy. For the case $L=10$, the charge conductance fluctuation
$G_{f}^{e}$ increases monotonously with the increasing of the
electron energy except a broad peak appears near the band edge to
the left, which corresponds to the peak in the density of states
(see reference \cite{Nikolic3}). However, the charge conductance
fluctuation for the case $L=40$ increases slowly with an increase of
the electron energy and its magnitude is very close to the universal
value of the charge conductance fluctuation $0.729e^{2}/h$
\cite{Lee3}. Interestingly, the charge conductance fluctuation
remains a constant value $0.5e^{2}/h$ in a wide regime of the
electron energy for the case $L=200$.   The spin conductance
fluctuation as function of the electron energy is plotted in figure
4(d). It is shown that the spin conductance fluctuation $G_{f}^{Sz}$
has the same sensitivity to the electron energy as that of the
average spin conductance $G_{a}^{Sz}$.

The ensemble average charge and spin conductances as function of the
wire length are illustrated in figures 5(a) and 5(b), respectively.
The electron energy $E$ is fixed at $2.08$. The average charge
conductance $G_{a}^{e}$ decreases sharply when the wire length is
smaller than the localization length $\lambda=140$. As the length of
the wire is increased further, the average charge conductance
decreases slowly. Conversely, when the wire length $L<140$, the average spin conductance
$G_{a}^{Sz}$ increases sharply with
increasing wire length and reach the maximum value $0.135e/4\pi$, indicating that the spin conductance comes from the
spin-dependent interference effect. However, the average spin conductance
decreases slowly as the wire length is increased further, since the charge conductance in this regime is
contributed by the resonance tunneling \cite{Bryant} or "necklace" states \cite{Pendry}. Figures
5(c) and 5(d) plot the charge and spin conductance fluctuations as
function of the wire length, respectively. The charge conductance
fluctuation $G_{f}^{e}$ also decreases sharply with an increase of the wire length except at
$L=3$, where a peak emerges. However, a constant value of the charge
conductance fluctuation $0.5e^{2}/h$ is found when the wire length is longer than the localization length $\lambda=140$.
The spin conductance fluctuation $G_{f}^{Sz}$ shows the
same sensitivity to the wire length as that of the average spin
conductance $G_{a}^{Sz}$.

\subsection{The conductance for a single Rashba wire with rough edges irradiated by a magnetic field}

Figures 6(a)-(c) show the charge conductance as function of the
electron energy for various magnetic field strengths. The strength
of the magnetic field in each panel is taken to be
$\hbar\omega_{c}=0, 0.5, 1.0$, respectively. The length of the
Rashba quantum wire is fixed at $L=40$. Many resonant peaks and
valleys exist in the charge conductance of the Rashba wire without
magnetic field, as shown in figure 6(a), resulting from the quantum
interference between the forward propagating channels and the
backward propagating channels caused by the rough boundaries.
However, those resonant conductance peaks and valleys are smeared
and the quantum charge conductance recover step by step with the
increasing of the magnetic field strength, as shown in figures 6(b)
and 6(c). These effects can be apprehended by the corresponding
energy band for the Rashba wire with the smallest width $W(x)=14$
shown in the inset of each figure, which determines the charge
conductance of the whole system. Each pair of the energy subbands
are lifted and the energy gap between two adjacent pair of subbands
is enlarged dramatically when the magnetic field strength is
increased, basically equaling $\hbar\omega_{c}$. Thus, the quantum interference induced by the edge
roughness is reduced, restoring the quantum charge conductance. The
corresponding spin conductance as function of the electron energy for each
magnetic field strength is plotted in figures 6(d)-(f),
respectively. Interestingly, the spin conductance disappears
gradually with an increase of the magnetic field strength. However,
a very large magnitude of the spin conductance ($\sim e^{2}/h$) can
be achieved between the two thresholds of each pair of subbands,
i.e., a energy window with a large spin conductance is opened.
Further, the width of the energy window increases with the raising
of the subbands. These phenomena can also be understood by the
energy band shown in each inset of figures 6(a)-(c).
Due to the energy gaps between two adjacent pair of subbands are
enlarged by the magnetic field, the Rashba subbands intermixing
becomes negligible \cite{Mireles} and the spin-dependent quantum
interference cause by the rough edges is reduced, resulting in the
disappearance of the spin conductance. In addition, the Rashba SOI
will lead to a Zeeman-like energy-band split when the
perpendicular magnetic field is placed to the Rashba wire and its energy
difference is $\Delta \varepsilon=(2\alpha\sqrt{2m^{\ast}}/\hbar)\sqrt{\hbar\omega_{c}(n+1)}$ \cite{Wang2}, where
$n$ denotes the index of the subband. Consequently,
the magnitude of the spin conductance inside the energy windows is
very large and the width of the energy window increases with the raise of the
subbands.

\section{Summary and conclusion}

In summary, we have clarified the spin-dependent electron transport
properties of a Rashba quantum wire with rough boundaries attached
two normal leads, and found that (i) a spin conductance can be
generated due to the broken structure symmetries and the
spin-dependent quantum interference caused by the rough edges; (ii)
the magnitude of the spin conductance will be enhanced firstly, and
then suppressed as the wire length is increased; (iii) the influence
of the rouge edges on the charge and spin conductances can be
eliminated when a perpendicular magnetic field is added to the
Rashba quantum wire. However, a very large amplitude of the spin
conductance can be generated when the electron energy is located
between the two thresholds of each pair of subbands. These
interesting findings may not only be useful in further understanding
the spin-dependent electron transport in low-dimensional Rashba
quantum structures but also provide some theoretical instructions in
the preparation of spintronics devices.

\section*{Acknowledgments}

This work was supported by the National Natural Science Foundation
of China (Grant No. 11147156, 10832005 and 10974052), and by the
Research Foundation of Jiangxi Education Department (Grant No.
GJJ12532).

\newpage
\section*{Figure captions}

~~~

Figure 1. (a) Schematic diagram of the Rashba quantum wire with
rough edges, connected to two ideal semi-infinite leads with a
vanishing SOI. The two leads have the same width $W$. The wire has
length $L$ and width $W(x)$, randomly fluctuating with $x$.

~~~

Figure 2. (Color online) (a) The calculated charge conductance as
function of the electron energy of a single sample of disordered
Rashba quantum wire with length: $L=10, 40$ and $200$. The dotted
line shows the charge conductance of a perfect Rashba wire of the
width $W(x)\equiv17$, length $L=10$, and leads $W=17$. (b) The
amplitude of the spin conductance as function of the electron energy
for the same wires. (c) The localization length for the Rashba wire
with boundary roughness. Graphs in (b) are vertically offset for
2.0.

~~~

Figure 3. (a) The calculated charge conductance as function of the
wire length of a single sample of disordered Rashba quantum wire.
(b) The amplitude of the spin conductance as function of the wire
length for the same wire. The electron energy $E=2.08$.

~~~

Figure 4. (Color online) The average charge (a) and spin (b)
conductances as function of the electron energy for different wire
lengths: $L=10, 40$ and $200$. The charge (c) and spin (d)
conductance fluctuations corresponding to the case in (a) and (b),
respectively. Number of samples taken for calculating average and
fluctuations value is 1000.

~~~

Figure 5. The average charge (a) and spin (b) conductances as
function of the wire length for the electron energy $E=2.08$. The
charge (c) and spin (d) conductance fluctuations corresponding to
the case in (a) and (b), respectively. Number of samples taken for
calculation is the same as that in figure 4.

~~~

Figure 6. (Color online) (a)-(c) The charge conductance as function
of the electron energy of a single disordered Rashba quantum wire
for different magnetic field strengths: $\hbar\omega_{c}=0,0.5,$ and
$1.0$. The inset in each panel is the energy band of a straight
Rashba wire with width $W(x)=14$, placed in a perpendicular magnetic
field. (d)-(f) The amplitude of the spin conductance as function of
the electron energy. The wire length is set at $L=40$.

\newpage

\begin{figure}
\center
\includegraphics[width=4.5in]{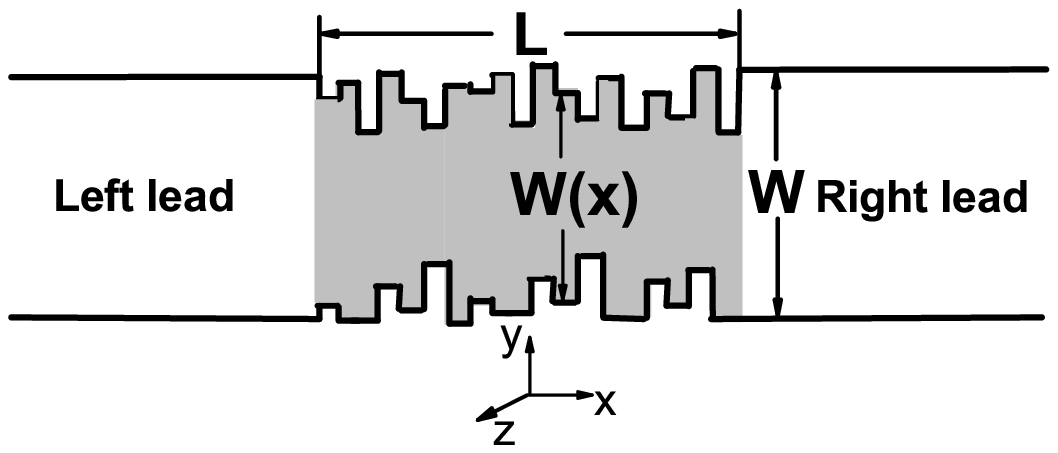}
\center{Figure 1}
\end{figure}

\begin{figure}
\center
\includegraphics[width=4.5in]{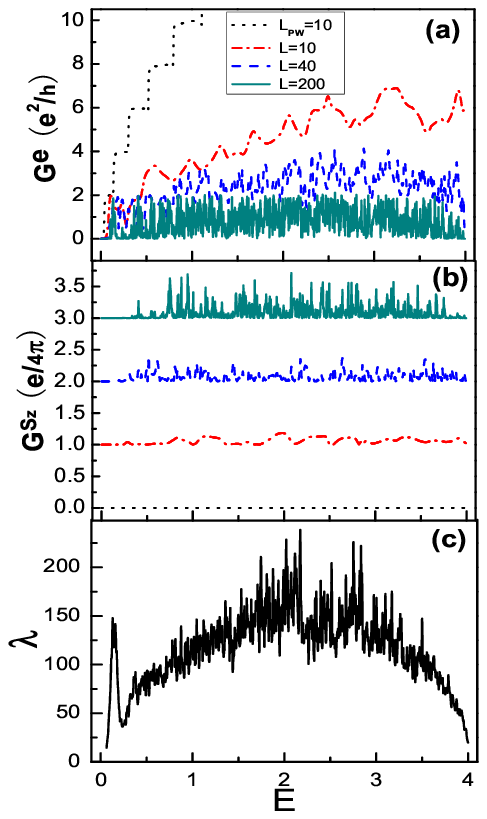}
\center{Figure 2}
\end{figure}

\begin{figure}
\center
\includegraphics[width=4.0in]{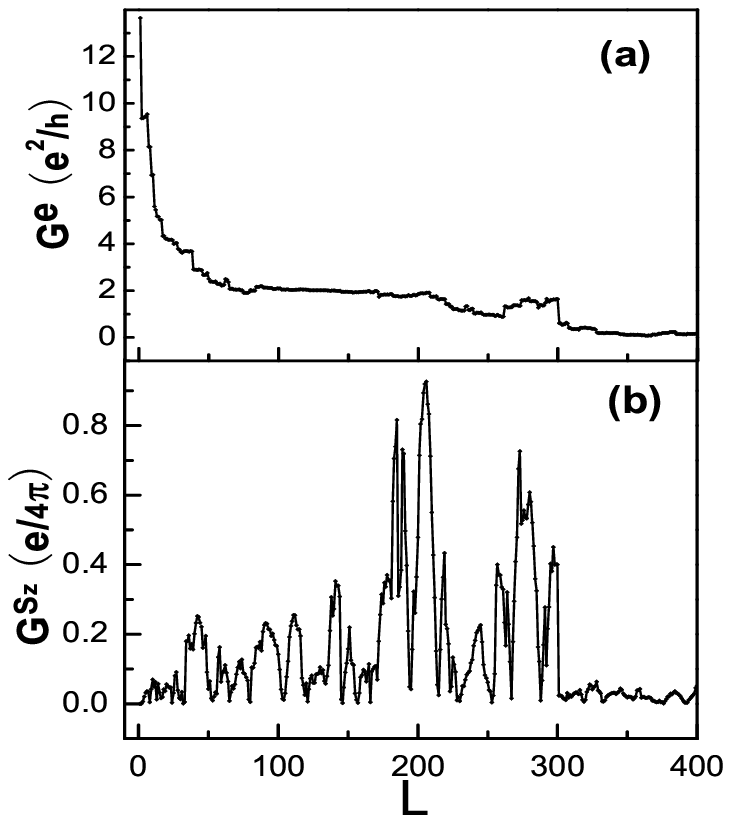}
\center{Figure 3}
\end{figure}

\begin{figure}
\center
\includegraphics[width=5.0in]{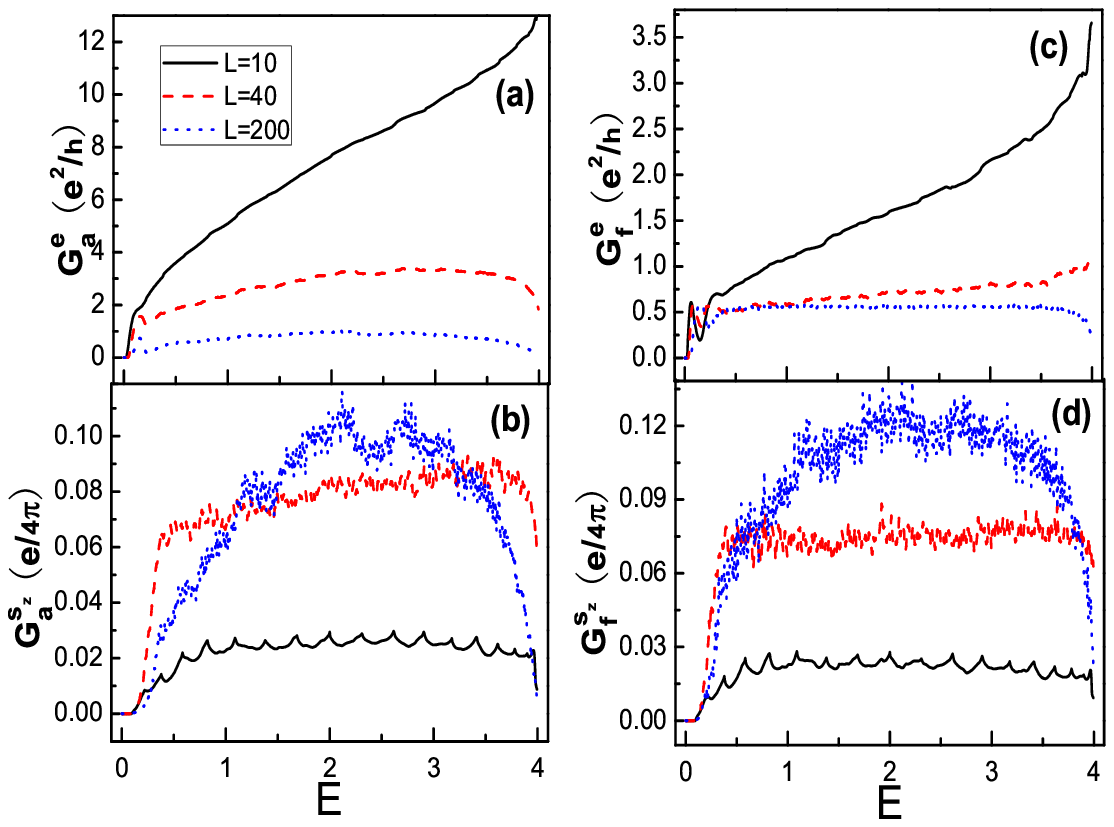}
\center{Figure 4}
\end{figure}

\begin{figure}
\center
\includegraphics[width=5in]{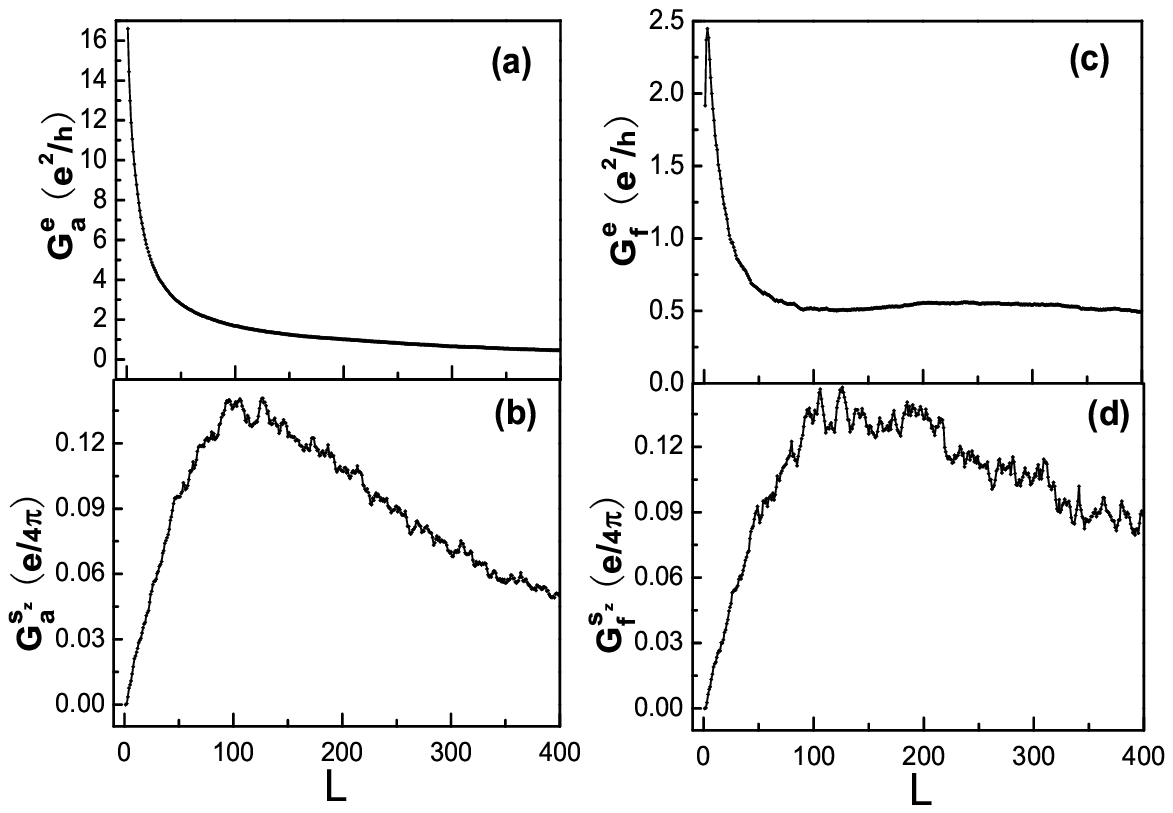}
\center{Figure 5}
\end{figure}

\begin{figure}
\center
\includegraphics[width=5in]{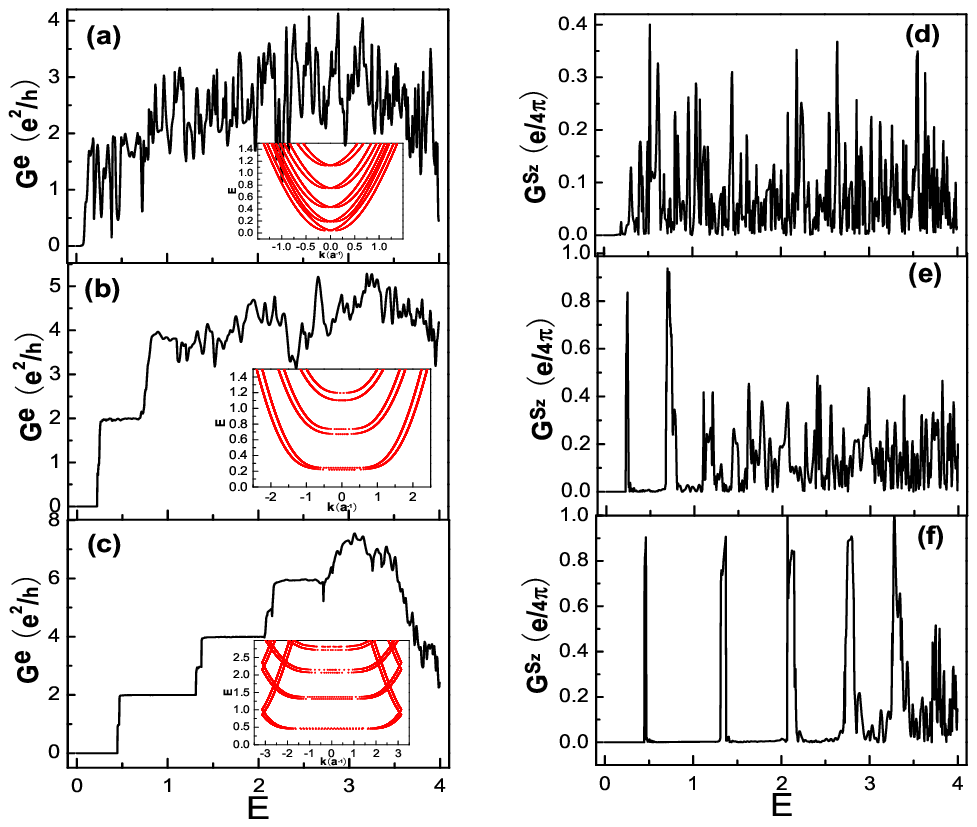}
\center{Figure 6}
\end{figure}

\end{document}